\documentclass[9pt,twocolumn,twoside]{opticajnl}
\journal{opticajournal} 

\setboolean{shortarticle}{false}

\usepackage{ulem}
\usepackage{lineno}

\title{Coherent ultraviolet frequency combs generated by resonant dispersive waves}

\author[1,2]{Michael Hemsworth}
\author[1,2]{Arthur K. Mills}
\author[1,2]{Christina Hofer}
\author[3]{TJ Hammond}
\author[1,2,*]{David J. Jones}

\affil[1]{Department of Physics and Astronomy, University of British Columbia, Vancouver, British Columbia, V6T 1Z1 Canada}
\affil[2]{Quantum Matter Institute, University of British Columbia, Vancouver, British Columbia, V6T 1Z4 Canada}
\affil[3]{Department of Physics,University of Windsor, Windsor, Ontario N9B 3P4 Canada}

\affil[*]{djjones@physics.ubc.ca}

\begin{abstract}
Resonant dispersive wave (RDW) emission in gas-filled anti-resonant hollow-core fibers (AR-HCF) offers a wavelength-tunable route to ultraviolet (UV) and vacuum ultraviolet (VUV) frequency comb generation. RDW emission is expected to be coherent fundamentally, but various noise sources are expected to limit the coherence, including amplitude-to-phase fluctuations due to input pulse intensity noise and plasma formation. It is critical, therefore, to establish the limits of the coherence of UV RDW pulses to determine their applicability for precision spectroscopy. Here, we demonstrate the coherence of RDW emission at 260~nm generated with an amplified Yb-doped fiber system at a 2-MHz repetition rate with a self-heterodyne beat measurement between the RDW and the fourth harmonic of the source. We isolate the RDW contribution to the frequency noise by coherently subtracting the carrier-envelope offset noise contribution measured simultaneously with an f-2f interferometer. We observe a beatnote linewidth below 10~kHz over timescales longer than 100~ms, and we find a conservative upper bound of 1~rad on the integrated RMS phase deviations over a frequency range spanning 1 to 30~kHz. The dominant frequency noise contributions are consistent with environmental noise of the interferometers and AR-HCF coupling fluctuations. These results provide strong evidence that RDW emission is coherent under well-controlled experimental conditions, confirming its potential for UV and VUV frequency comb generation.

\end{abstract}

\setboolean{displaycopyright}{false} 

\begin{document}

\maketitle

\section{Introduction} \label{sec:intro}

Demand is increasing for ultrafast light sources in the ultraviolet (UV), vacuum ultraviolet (VUV), and extreme ultraviolet (XUV) spectral regions for applications in dual-comb spectroscopy \cite{mccauleyDualcombSpectroscopyDeep2024a,xuNearultravioletPhotoncountingDualcomb2024,schusterUltravioletDualComb2021,kirchnerUltrabroadbandUVVIS2025}, photoemission spectroscopy \cite{brombergerAngleresolvedPhotoemissionSpectroscopy2015a,kotsinaUltrafastMolecularSpectroscopy2019,kotsinaSpectroscopicApplicationFewfemtosecond2022,naAdvancingTimeAngleresolved2023},  and precision spectroscopy with femtosecond frequency combs (FFCs) \cite{schonbergBelowthresholdHarmonicGeneration2023, cingozDirectFrequencyComb2012, witteDeepUltravioletQuantumInterference2005, morgenwegRamseycombSpectroscopyIntense2014}. In all cases, frequency conversion with multiple second-order nonlinearities or high-order nonlinearities is required to translate the FFCs (typically in the near-infrared) to the UV spectral range. A trade-off frequently emerges: nonlinear conversion favors high peak intensity, while the experimental signal-to-noise ratio favors high repetition rates. At fixed average power, adjusting the repetition rate shifts the balance in favor of one at the expense of the other \cite{canellaLowrepetitionrateOpticalFrequency2024a}. The technical demands of this trade-off place a premium on new sources that provide broad wavelength tunability via efficient nonlinear conversion at spectroscopically useful repetition rates, while preserving the coherence of the FFC.

Two complementary approaches to precision spectroscopy with such combs are frequency-domain techniques such as direct frequency comb spectroscopy (DFCS) \cite{marianDirectFrequencyComb2005, cingozDirectFrequencyComb2012} or dual-comb spectroscopy \cite{mccauleyVacuumUltravioletDualComb2026}, where the frequency noise and strength of individual comb lines are critical, and time-domain techniques such as Ramsey-comb spectroscopy (RCS), where the relative phase of amplified comb pulses separated by a variable delay carries the spectroscopic information \cite{morgenwegRamseycombSpectroscopyIntense2014}. In both approaches, the resolution scales inversely with the coherent interrogation time, although RCS additionally requires coherent amplified pulse pairs at increasingly long delays. In the VUV and XUV, high-repetition-rate combs generated through high-order harmonic generation have enabled DFCS \cite{cingozDirectFrequencyComb2012}, demonstrated coherence times exceeding one second \cite{benkoExtremeUltravioletRadiation2014}, and recently allowed direct comb spectroscopy of the $^{229}$Th nuclear transition \cite{zhangFrequencyRatio229mTh2024}. More recently, the properties of low-repetition-rate FFCs for DFCS has been investigated \cite{canellaLowrepetitionrateOpticalFrequency2024a}. RCS has achieved $10^{-11}$-scale relative frequency uncertainties in the deep UV using microsecond pulse delays \cite{altmannHighPrecisionRamseyCombSpectroscopy2016,altmannDeepUltravioletFrequencyMetrology2018}.  

Nonlinear conversion in crystals can preserve comb coherence, but transparency and phase-matching constraints generally restrict operation to wavelengths above approximately 190~nm, with specialized crystals extending toward 150~nm \cite{nakazatoPhasematchedFrequencyConversion2016a}. Gas-phase four-wave mixing and high-order harmonic generation reach shorter wavelengths, although they typically require either low-repetition-rate amplified systems or enhancement cavities, and HHG produces only discrete harmonics of the driving laser \cite{ghotbiGeneration25MJ2010,benkoExtremeUltravioletRadiation2014,millsCavityenhancedHighHarmonic2019}. Gas-filled hollow-core fibers substantially reduce the required pulse energy by confining the optical field over an extended interaction length. This approach has enabled efficient four-wave mixing into the VUV using both large-core fibers at repetition rates approaching 100~kHz \cite{forbesEfficient10Generation2024,forbesHighlyEfficientDUV2025} and small-core anti-resonant hollow-core fibers (AR-HCFs) at kHz \cite{belliHighlyEfficientDeep2019} and MHz \cite{couchUltrafast1MHz2020b} repetition rates.  

Of particular interest here is resonant dispersive wave (RDW) emission in gas-filled AR-HCFs. The emission wavelength can be tuned through the gas species, pressure, fiber geometry, and driving conditions, while pump-to-UV conversion efficiencies exceeding 5\% have been demonstrated \cite{kottigGenerationMicrojoulePulses2017}. These characteristics provide an attractive and flexible means of targeting the many atomic and molecular transitions distributed throughout the UV and VUV. During soliton self-compression, energy can be transferred efficiently from the driving pulse into the phase-matched RDW band. RDW emission has been demonstrated down to approximately 115~nm in kilohertz systems \cite{ermolovSupercontinuumGenerationVacuum2015}, while microjoule-level UV pulses with conversion efficiencies of several percent have been generated at repetition rates approaching 2~MHz \cite{kottigGenerationMicrojoulePulses2017}. RDW emission therefore combines efficient conversion with broad tunability, providing access to spectral regions not readily reached by crystal conversion or fixed-order harmonic generation. These properties make it a promising route to UV and VUV FFCs, provided that the coherence of the driving comb is preserved during the strongly nonlinear conversion process.

Although RDW emission is a deterministic, phase-matched process, its strong nonlinear dependence on the driving pulse provides several pathways through which excess frequency noise could arise. Simulations predict high first-order coherence across RDW spectra \cite{dudleyCoherencePropertiesSupercontinuum2002,chen_octave-wide_2024}, but also show that driving-pulse energy fluctuations can be transferred to the energy and arrival time of the RDW, particularly below saturation \cite{brahmsTimingEnergyStability2021}. This amplitude-to-phase coupling could degrade the RDW coherence. Photoionization provides another possible pathway because intensity-dependent plasma formation modifies the soliton dynamics and the RDW phase-matching condition \cite{chang_influence_2011,saleh_theory_2011}. At high repetition rates, long-lived refractive-index and gas-density perturbations may also affect subsequent pulses \cite{koehlerLongLivedRefractiveIndexChanges2018,tani_long-lived_2025}. These studies identify plausible mechanisms for decoherence, but do not directly measure their effect on the pulse-to-pulse optical phase of the emitted RDW.

Ermolov et al. performed a phase-sensitive characterization of ultraviolet RDW emission by interfering an RDW with the second harmonic of the driving pulse near 415~nm \cite{ermolovCarrierenvelopephasestableSolitonbasedPulse2019}. The observation of stable interference fringes demonstrates that, in this case, the visible RDW retains a definite phase relationship to the driving field. They measured approximately 300~mrad RMS carrier-envelope-phase variation over 0.5–240~Hz and observed features suggesting coupling between pump intensity and RDW phase. Kirchner et al. investigated comb preservation using an f-2f interferometer and measured an approximately 100-kHz-wide $f_0$ beat at 515~nm \cite{kirchnerUltrabroadbandUVVIS2025}. The measurement was limited by the natural carrier-envelope offset frequency linewidth of the unstabilized oscillator and was performed in the green supercontinuum tail rather than at the 330~nm RDW wavelength. It therefore provided only indirect evidence for coherence of the RDW itself.

In this work, we directly characterize coherence at the RDW wavelength using an f-4f interferometer to generate a heterodyne beat between the UV RDW and the fourth harmonic of the driving laser. This beat retains the oscillator's carrier-envelope offset frequency noise together with excess noise introduced during the RDW generation. A simultaneous f-2f measurement is used to remove the oscillator's carrier-envelope offset frequency contribution.  We measure less than 1~rad RMS excess phase jitter integrated from 1 to 30~kHz, providing strong evidence that RDW emission preserves the phase coherence of the driving frequency comb. Note that as this result also includes (confounding) noise introduced by the interferometer's path fluctuations, it establishes an upper bound on the excess RDW phase noise.

\section{Coherence Measurement Principles}

As mentioned previously, we utilize an f-4f interferometer (shown as a component in Fig. \ref{fig:setup}) to quantify a possible loss of coherence in the RDW emission process. In this case, the 4f-arm frequency quadruples the driving FFC, while the f-arm includes the AR-HCF that generates a FFC in the UV via RDW emission. When the two arms are spectrally and spatio-temporally overlapped, their interference produces an optical heterodyne beat containing the carrier-envelope offset frequency of the driving oscillator, together with frequency noise contributions from the interferometer and the RDW emission process. Specifically, the beat frequency can be expressed as
\begin{equation}
\label{eq:f_f4f}
f_{f-4f}(t) = K' f_{\mathrm{rep}}(t) + 3f_0(t) - f_{\mathrm{RDW}}(t) + f_{\mathrm{int}}'(t),
\end{equation}
where $f_{\mathrm{rep}}(t)$ is the repetition rate, $K'$ is an integer specifying the comb tooth numbers involved in the mixing process, $f_0(t)$ is the carrier-envelope offset frequency and $f_{\mathrm{int}}'(t)$ accounts for interferometer-instability-induced noise in the f-4f configuration. The $f_{f-4f}(t)$ beat has a term proportional to $3f_0(t)$ and $f_{\mathrm{RDW}}(t)$ represents the additional frequency noise imparted by the RDW generation process and is the primary quantity of interest in this work. Note that we have explicitly added time dependence to each term to emphasize that both the comb parameters and the various other contributions each have their own associated frequency noise.

The oscillator is not equipped with active stabilization of the offset frequency $f_0(t)$. Consequently, fluctuations in $f_0(t)$ contribute directly to the measured f-4f beat and need to be separated from the contribution of $f_{\mathrm{RDW}}(t)$. We therefore simultaneously measure the f-2f beat of the driving oscillator, providing an independent measurement of $f_0(t)$ that can be used to remove the $3f_0(t)$ contribution in Eq.~\ref{eq:f_f4f}. The instantaneous f-2f beat frequency can be expressed as
\begin{equation}
\label{eq:f_f2f}
f_{f-2f}(t) = K f_{\mathrm{rep}}(t) + f_0(t) - f_{\mathrm{NL}}(t) + f_{\mathrm{int}}(t),
\end{equation}
where $K$ is an integer that specifies the particular pair of comb teeth involved in the mixing process. The term $f_{\mathrm{NL}}(t)$ represents frequency noise imparted by the nonlinear supercontinuum generation process, while $f_{\mathrm{int}}(t)$ accounts for additional frequency noise introduced by interferometer instabilities.

By combining Eqs.~\ref{eq:f_f4f} and \ref{eq:f_f2f}, we can construct a difference-frequency signal that removes the common-mode carrier-envelope offset frequency $f_0(t)$. Specifically, subtracting three times the f-2f beat note from the f-4f beat note yields
\begin{equation}
\label{eq:f_diff}
\begin{split}
f_{\mathrm{diff}}(t) &= f_{f-4f}(t) - 3f_{f-2f}(t) \\
&= \Delta K f_{\mathrm{rep}}(t) + \Delta f_{\mathrm{int}}(t) + 3f_{\mathrm{NL}}(t) - f_{\mathrm{RDW}}(t),
\end{split}
\end{equation}
where $\Delta K = K' - 3K$ and $\Delta f_{\mathrm{int}}(t) = f_{\mathrm{int}}'(t) - 3f_{\mathrm{int}}(t)$. In this expression, the $f_0(t)$ contribution is fully cancelled, leaving a signal that is sensitive to differential noise sources between the two interferometers. While Eq.~\ref{eq:f_diff} includes the desired $f_{\mathrm{RDW}}(t)$ contribution, it also contains residual nonlinear and interferometer noise, such that the measured fluctuations in $f_{\mathrm{diff}}(t)$ provide an upper bound on the noise imparted by the RDW generation process. A more detailed derivation of Eqs.~\ref{eq:f_f4f}, \ref{eq:f_f2f}, and \ref{eq:f_diff} is provided in Supplement~\ref{sec:supp_s1}.

\section{Experimental setup} 

\begin{figure}[ht]
    \centering
    \includegraphics[width=\linewidth]{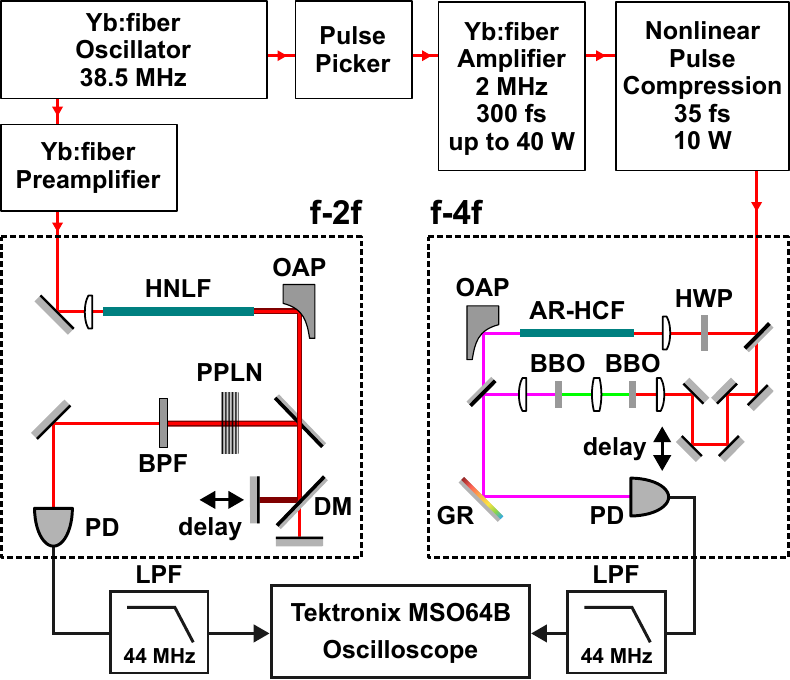}
    \caption{Experimental setup for the simultaneous f-2f and f-4f beat note measurements. The f-4f interferometer compares ultraviolet resonant dispersive wave emission from an anti-resonant hollow-core fibers with the fourth harmonic of the driving femtosecond frequency comb, while the independent f-2f interferometer monitors the oscillator $f_0$. HNLF, highly nonlinear fiber; OAP, off-axis parabolic mirror; PPLN, periodically poled Lithium Niobate; BPF, bandpass filter; PD, photodiode; DM, dichroic mirror; LPF, lowpass filter; HWP, half-wave plate; GR, grating. }
    \label{fig:setup}
\end{figure}

The experimental setup is shown in Fig.~\ref{fig:setup}. A Yb-doped fiber oscillator based on the design of Ref.~\cite{mayerFlexibleAllPMNALM2020a} is modified for operation at 38.5~MHz and is centered at 1035~nm. The majority of the oscillator output power is sent to a high-power, pulse-picked chirped-pulse amplifier for subsequent RDW generation and f-4f interferometry (lower-right dashed box). A small portion of the oscillator output is diverted to a single-mode Yb-doped fiber preamplifier that seeds an f-2f interferometer used to monitor the oscillator $f_0$ (lower-left dashed box). The oscillator $f_0$ is not actively stabilized.

To reach sufficient pulse energies for RDW UV generation, pulse picking down to 2~MHz is implemented using an acousto-optic modulator (AOM). The AOM carrier frequency is derived from the eighth harmonic of the oscillator repetition rate to preserve intrinsic carrier-envelope phase stability through the pulse picking operation \cite{devriesAcoustoopticPulsePicking2015}. Although pulse picking modifies the comb structure, the resulting beat note expressions retain the same functional dependence on $f_{\mathrm{rep}}(t)$ and $f_0(t)$. The $f_0(t)$ cancellation arguments presented above therefore remain valid (For details see Supplement~\ref{sec:supp_s1}). Subsequent chirped-pulse amplification produced a pulse train with a pulse duration of approximately 300~fs. As RDW emission requires shorter pulses, an AR-HCF-based spectral broadening stage followed by chirped mirror compression generates pulses with a 35~fs duration \cite{hemsworth2026Addressing}.

The compressed pulse train is split into the two arms of the f-4f interferometer using a half-wave plate and a pair of thin-film polarizers that allow for power control (not shown). In the upper arm, an additional half-wave plate rotates the beam to p-polarization to match the polarization in the lower arm. An $f=75$~mm plano-convex lens then focuses the beam into a 20-cm-long AR-HCF with a 30~\textmu{}m core diameter and a capillary wall thickness of 300~nm. The fiber is housed in a gas cell filled with approximately 11~bar argon. The resulting RDW light at 260~nm is recollimated using a UV-enhanced aluminum off-axis parabolic mirror (OAP) with a focal length of 100~mm and directed toward a 50/50 UV beamsplitter. The lower arm first passes through a manual delay line used to achieve temporal overlap. An $f=50$~mm lens focuses the beam into a 200-\textmu{}m-thick beta-barium borate (BBO) crystal cut for second-harmonic generation at 1040~nm. An $f=25$~mm biconvex lens then refocuses the generated light into a second 200-\textmu{}m-thick BBO crystal cut for second-harmonic generation at 520~nm. The resulting 260~nm light is collimated by an $f=75$~mm plano-convex lens and combined with the upper arm at the 50/50 UV beamsplitter. The combined beams are spectrally dispersed by a grating, and a slit selects the overlapping region near 260~nm before the light reaches a UV photodiode (PD) that detects the heterodyne beat.

In the f-2f interferometer, light is focused by an objective lens into a highly nonlinear photonic crystal fiber (HNLF) for octave-spanning supercontinuum generation. After recollimation with an OAP, the f and 2f portions of the spectrum are separated with a dichroic mirror (DM) into two arms with adjustable delay. Both arms are then recombined on the same DM with a slight vertical tilt and picked off with a d-mirror. A periodically poled lithium niobate (PPLN) crystal doubles the f component to spectrally overlap with the 2f component. A bandpass filter (BPF) selects the overlapping spectral region, and the two components interfere on a PD to generate the heterodyne beat signal. The f-2f and f-4f PD signals are lowpass filtered at 44~MHz and recorded simultaneously with a Tektronix MSO64B oscilloscope at a sampling rate of 125~MS/s and a record length of 25~MS, corresponding to a total measurement time of 0.2~s. The time series data are then post-processed to extract the uncorrelated noise contributions.

\begin{figure}[ht]
    \centering
    \includegraphics[width=0.9\linewidth]{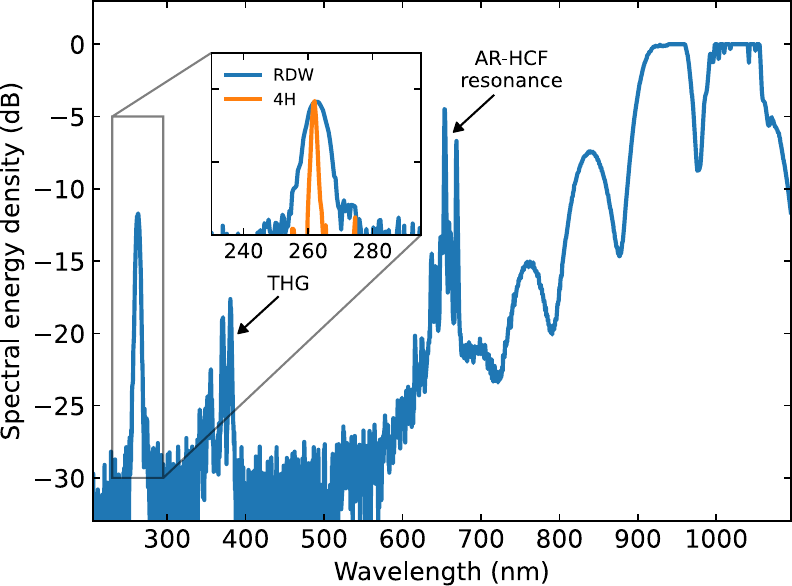}
    \caption{Measured optical spectrum of the RDW fiber output. Part of the supercontinuum near 1000~nm is flattened due to spectrometer saturation. The long-wavelength edge above 1100~nm is limited by the Si detector response of the spectrometer. The inset shows a zoom-in of the RDW spectrum with a spectrum of the fourth harmonic (4H) arm overlayed. Spectral features attributed to third-harmonic generation (THG) and phase matching near AR-HCF wall resonances are indicated.}
    \label{fig:rdw_4hg_spectrum}
\end{figure}

Figure~\ref{fig:rdw_4hg_spectrum} shows the output optical spectrum from the AR-HCF arm of the f-4f interferometer. The spectrum consists of a broad supercontinuum spanning approximately 700--1100~nm and three narrower spectral features. The detected long-wavelength edge is limited by the Si detector response, while part of the supercontinuum near 1000~nm is clipped by spectrometer saturation. The feature near 650~nm is attributed to a parasitic resonant process phase-matched by capillary-wall resonances in the AR-HCF \cite{taniEffectAnticrossingsCladding2018}, while the feature near 370~nm is attributed to non-phase-matched third-harmonic generation (THG). The feature near 260~nm corresponds to the RDW emission used in this work. The inset compares a zoomed-in RDW spectrum with an independently measured spectrum from the fourth harmonic (4H) generation arm, confirming the spectral overlap between the two arms of the f-4f interferometer.

\begin{figure}[ht]
    \centering
    \includegraphics[width=\linewidth]{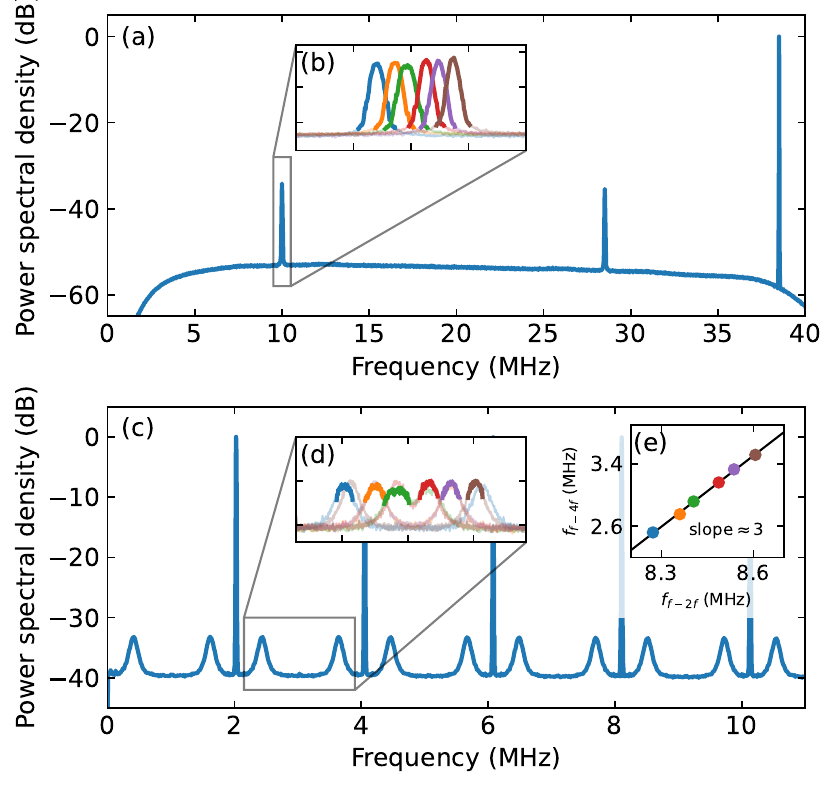}
    \caption{Power spectral densities of the measured f-2f and f-4f beat notes in panels (a) and (c) respectively. The expanded views in panels (b) and (d) show six measurements obtained at different oscillator pump currents. Because the complementary f-4f beat notes are closely spaced, the peaks whose shifts are correlated with the positive frequency shift of the f-2f beat note are highlighted. The inset (e) displays the first moment frequencies of the corresponding highlighted f-2f and f-4f beat notes in (b) and (d). A linear fit with a slope of approximately 3 demonstrates the $3f_0$ scaling of the f-4f beat note.}
    \label{fig:rf_spectra}
\end{figure}

Power spectral densities (PSDs) of the f-2f and f-4f beat notes calculated from time series data using Welch's method \cite{welchUseFastFourier1967} are shown in Fig.~\ref{fig:rf_spectra}(a) and (c), respectively. The f-2f beat notes have a signal-to-noise ratio (SNR) of approximately 20~dB, whereas the f-4f beat notes lie only approximately 5~dB above the noise floor. We did not use balanced detection in either interferometer which may improve the observed SNR \cite{ruehlUltraBroadbandCoherentSupercontinuum2011}.

According to Eqs.~\ref{eq:f_f4f} and \ref{eq:f_f2f}, the f-4f beat note is expected to be proportional to $3f_0(t)$, whereas the f-2f beat note has the usual $f_0(t)$ dependence. Consequently, a change in $f_0(t)$ is expected to produce a frequency shift in the f-4f beat that is approximately three times the corresponding shift in the f-2f beat. We verify this dependence by recording six pairs of f-2f and f-4f beat notes while making small adjustments of $f_0(t)$ (via the oscillator pump current). The individual beat note spectra obtained at each pump current are shown in different colours in panels (b) and (d) of Fig.~\ref{fig:rf_spectra}. The corresponding first-moment frequencies of the f-2f and f-4f beat notes are plotted against each other in Fig.~\ref{fig:rf_spectra}(e). A linear fit gives a slope of approximately 3, in agreement with the expected $f_0(t)$ and $3f_0(t)$ dependence of the f-2f and f-4f beat notes, respectively.

\section{RDW noise extraction}

\begin{figure*}[t]
    \centering
    \includegraphics[width=0.8 \linewidth]{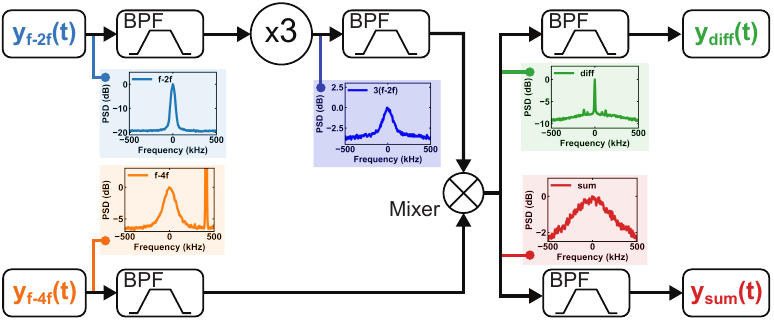}
    \caption{Digital signal processing steps for beat note mixing. The time-domain f-2f signal, $y_{f-2f}(t)$, is bandpass filtered (BPF) in the frequency domain with a passband centered at $f_{f-2f}$ (light blue). The filtered signal is then cubed in the time domain to generate a frequency-tripled component (x3) and bandpass filtered again at $3f_{f-2f}$ (dark blue). The f-4f signal, $y_{f-4f}(t)$, is bandpass filtered with a passband centered at $f_{f-4f}$ (orange). The two filtered signals are multiplied in the time domain to produce sum- and difference-frequency components (red and green, respectively), and a final bandpass filter isolates the desired component. The beat note power spectral densities (PSDs) at each stage are also shown for reference.}
    \label{fig:mixing_schematic}
\end{figure*}

The reduced signal-to-noise ratio and the tight frequency spacing of the RDW comb makes an analog-based approach to isolating $f_{\mathrm{RDW}}(t)$ technically challenging so we turn to digital signal processing. These post-processing steps applied to the measured f-2f and f-4f time-domain signals, $y_{f-2f}(t)$ and $y_{f-4f}(t)$, are illustrated in Fig.~\ref{fig:mixing_schematic}. A single beat note from each interferometer is first isolated using a frequency-domain BPF. The f-2f signal is then cubed in the time domain to generate a component with the required $3f_0(t)$ scaling. A second bandpass filter isolates this tripled component and suppresses the remaining components generated by the frequency multiplication. The filtered f-4f and tripled f-2f signals are subsequently multiplied in the time domain, producing sum- and difference-frequency components. A final bandpass filter isolates the difference-frequency component. In this signal, the correlated $3f_0(t)$ fluctuations cancel, whereas in the sum-frequency signal they add to produce a $6f_0(t)$ contribution. The sum-frequency component is therefore examined as a consistency check and is expected to be substantially broader. The difference signal retains noise introduced during RDW generation together with uncorrelated contributions from nonlinear broadening in the f-2f interferometer and instabilities from both interferometers. It therefore provides an upper bound on the excess noise associated with RDW generation.

From the time-domain signals obtained at different stages of this procedure, we extract both the beat note PSD and the associated phase and frequency noise PSDs. The instantaneous phase is retrieved from each bandpass-filtered time series $y(t)$ using
\begin{equation}
\label{eq:inst_phase}
\phi(t) = \textup{unwrap}\big(\arg[y(t) + i\mathcal{H}\{y(t)\}]\big),
\end{equation}
where $\mathcal{H}\{\cdot\}$ denotes the Hilbert transform and the unwrap function removes $2\pi$ discontinuities. The residual phase fluctuation is defined as
\begin{equation}
\Delta\phi(t)=\phi(t)-(2\pi f_c t+\phi_0),
\end{equation}
where $2\pi f_c t+\phi_0$ is the best-fit linear trend associated with the carrier frequency $f_c$ of the beat signal with phase offset $\phi_0$. The phase noise spectrum, $S_{\Delta\phi}(f)$, is then obtained by computing the PSD of $\Delta\phi(t)$. The corresponding frequency noise spectrum is given by $S_{\Delta f}(f)=f^2S_{\Delta\phi}(f)$.

\section{Results/discussion}

\begin{figure}[ht]
    \centering
    \includegraphics[width=0.9\linewidth]{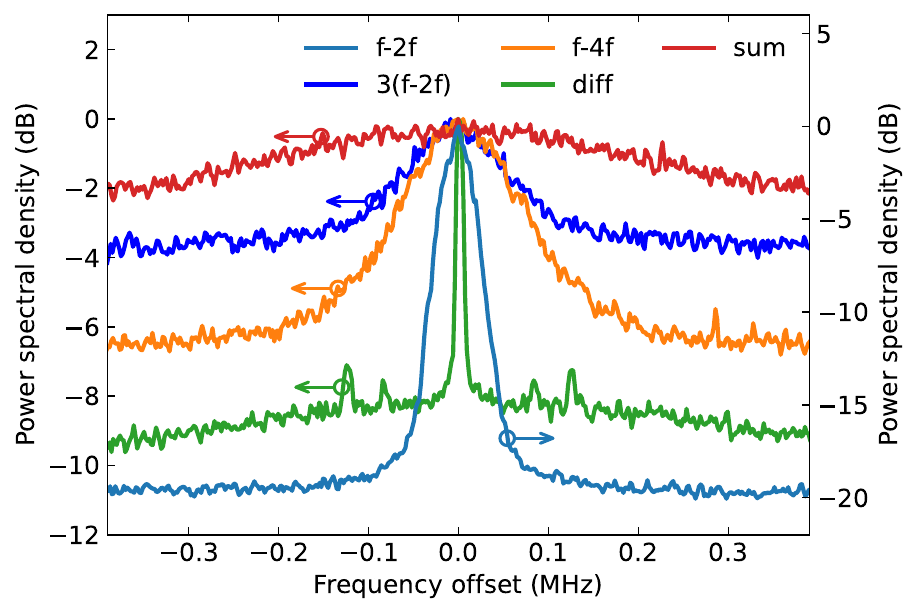}
    \caption{Beat note power spectral densities at successive stages of the signal-processing chain: f-2f (light blue), frequency-tripled f-2f (dark blue), f-4f (orange), mixed difference-frequency (green), and mixed sum-frequency (red) beat signals. Each spectrum is shifted to zero frequency to facilitate comparison of the relative lineshapes. The f-2f spectrum is plotted against the right-hand y-axis to improve the dynamic range of the other spectra. The narrow difference-frequency signal results from cancellation of the correlated $f_0(t)$ fluctuations, whereas these fluctuations add in the broader sum-frequency signal.}
    \label{fig:beat_notes}
\end{figure}

We first analyze the PSDs of the various beat signals at successive stages of the processing chain. Figure~\ref{fig:beat_notes} shows the PSDs of the f-2f (light blue), frequency-tripled f-2f (dark blue), f-4f (orange), mixed difference-frequency (green), and mixed sum-frequency (red) signals. Each beat is shifted to zero frequency to facilitate comparison of the relative lineshapes. The f-2f spectrum is plotted against the right-hand y-axis with a larger range to better illustrate fine details in the other spectra. After frequency tripling, the f-2f beat note more closely resembles the f-4f beat note in spectral shape, consistent with the expected $3f_0(t)$ scaling. Following the mixing procedure, the difference-frequency (sum-frequency) component exhibits a substantially narrower (broader) spectral peak than either input beat note, consistent with the correlated $f_0(t)$ fluctuations cancelling (adding) in the difference (sum) signal. The contrast between these components therefore demonstrates that the tripled f-2f and f-4f beat notes contain strongly correlated common-mode noise. The residual width of the difference-frequency signal includes all uncorrelated noise contributions and therefore establishes an upper bound on the excess noise introduced during RDW generation.

\begin{figure}
    \centering
    \includegraphics[width=0.9\linewidth]{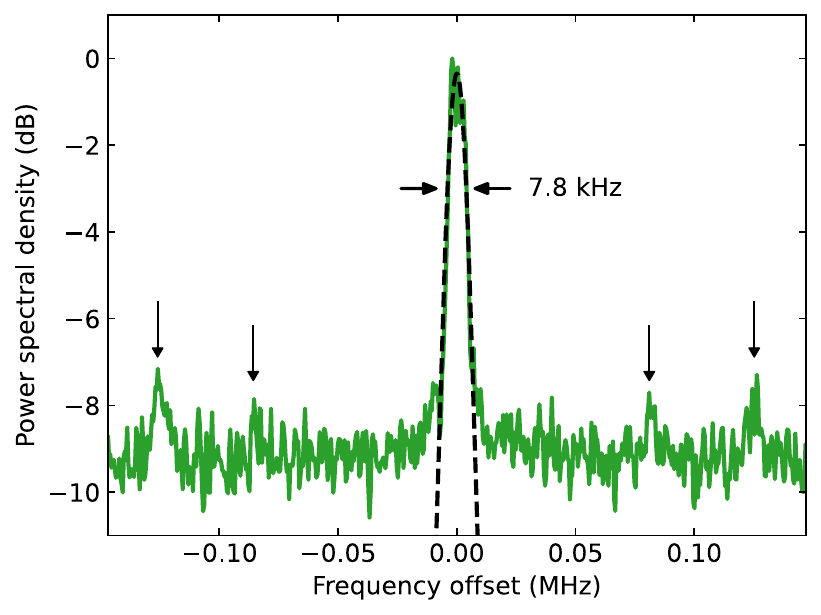}
    \caption{Beat note power spectral density of the isolated mixed difference-frequency component. The central peak has a FWHM linewidth of 7.8~kHz. The spectrum was obtained over a measurement period of 200~ms with an effective resolution bandwidth of 715~Hz.}
    \label{fig:diff_beat}
\end{figure}

To quantify this residual linewidth, Fig.~\ref{fig:diff_beat} shows the PSD of the isolated difference-frequency component. The central peak has a full width at half-maximum (FWHM) linewidth of 7.8~kHz over the 200~ms measurement period. The effective resolution bandwidth, calculated from the equivalent noise bandwidth of the spectral window, is 715~Hz. Its finite width characterizes the residual differential fluctuations remaining after cancellation of the correlated $f_0(t)$ noise. We trace the additional peaks at frequency offsets of 85 and 125~kHz (indicated by the vertical arrows) to parasitic modulation introduced by the pulse-picking electronics, as identical sidebands are observed in the RF spectrum of the electrical gate signal that controls the AOM.

To gain further insight into the origin of the observed linewidth broadening, we analyze the phase and frequency noise PSDs, computed from the time-dependent phase extracted from the beat note signals. This approach allows us to distinguish between different noise contributions and assess their relative impact on the spectral linewidth. The extraction of frequency noise is sensitive to the bandwidth of the BPFs used to isolate the individual beat notes. We must limit the filter width to 3-5 times the beat note FWHM to produce reliable phase extractions. To validate the selected filter bandwidths, we reconstruct the beat note spectrum lineshape from the extracted frequency noise and compare it to the beat note PSD following the approach of Ref.~\cite{didomenicoSimpleApproachRelation2010}. This comparison largely confirms that the filtering process does not distort the underlying noise characteristics within the filter bandwidth, which we can interpret as the upper frequency limit of the extracted frequency noise spectra. For further details see Supplement~\ref{sec:supp_s2}.

\begin{figure}[ht]
\centering
\includegraphics[width=\linewidth]{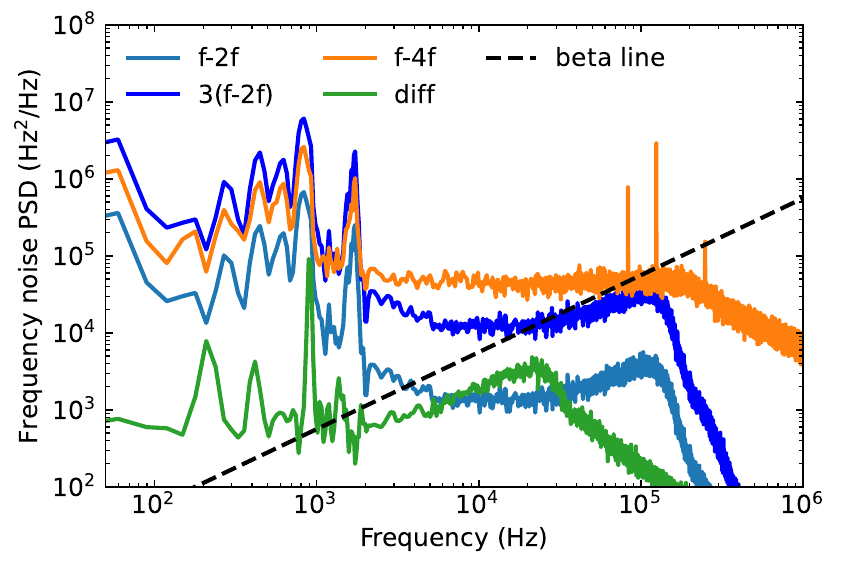}
\caption{Frequency noise power spectral density of the f-2f, tripled f-2f, f-4f and difference frequency mixed beat notes. The black dashed line represents the beta separation line.}
\label{fig:freq_noise}
\end{figure}

The frequency noise PSD of the f-2f, tripled f-2f, f-4f, and difference frequency beat signals are shown in Fig.~\ref{fig:freq_noise}. The black dashed line indicates the beta separation line, which provides a useful distinction between noise contributions that predominantly contribute to the linewidth (above the line) and those that primarily contribute to the spectral wings (below the line) \cite{didomenicoSimpleApproachRelation2010}. All spectra exhibit pronounced structure at frequencies below $\sim$2~kHz, while the higher-frequency region is relatively flat below the rolloff imposed by the finite width of the bandpass filters. Most notably, the tripled f-2f and f-4f spectra display similar peaks, indicating the presence of common-mode noise, likely originating from fluctuations in the oscillator carrier-envelope offset frequency, itself. This common-mode noise picture is corroborated by the frequency noise spectrum of the difference frequency signal, which shows a clear reduction in noise power across the entire frequency range. In particular, many of the low-frequency features present in the individual beat notes are strongly suppressed. The remaining structure is dominated by a prominent peak near $\sim$900~Hz, along with smaller peaks at approximately 200~Hz and 400~Hz. These residual components represent differential-mode noise that is not shared between the two interferometers. 

\begin{figure}[ht]
    \centering
    \includegraphics[width=\linewidth]{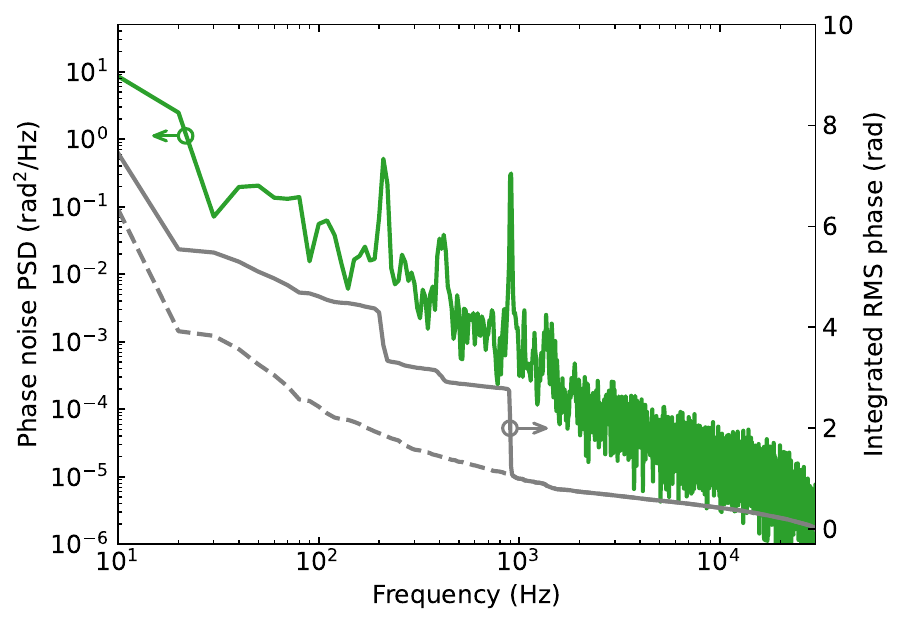}
    \caption{Phase noise power spectral density (green) and integrated RMS phase (gray) of the difference frequency beat signal. The gray dashed line represents the integrated RMS phase noise with the three peaks at 200, 400 and 900~Hz removed.}
    \label{fig:phase_noise}
\end{figure}

The phase noise PSD of the mixed difference-frequency beat note is shown in green in Fig.~\ref{fig:phase_noise}, together with the integrated RMS phase shown by the solid gray curve. Integrating the phase noise spectrum from 30~kHz down to 10~Hz yields a total RMS phase deviation of 7.45~rad. The dominant contributions arise from discrete peaks near 200, 400, and 900~Hz, as well as from the low-frequency region below approximately 20~Hz, while the higher-frequency portion of the spectrum contributes relatively little. To illustrate the contribution of the discrete spectral features, the dashed gray curve shows the integrated RMS phase recalculated after excluding narrow frequency intervals containing the peaks at 200, 400, and 900~Hz. This curve is included as a diagnostic comparison rather than as a corrected measurement of the intrinsic RDW phase noise. The measured integrated phase remains below 1~rad for lower integration limits down to approximately 1~kHz, corresponding to a coherence time of approximately 1~ms.

For comparison, fiber-based optical frequency combs operating in a phase-coherent regime can  exhibit integrated RMS phase noise of 250 mrad (\textmu{}Hz to Nyquist)   \cite{liehl_ultrabroadband_2017}. However, the larger value observed here (over a smaller frequency range) should  be interpreted as a conservative upper bound rather than an intrinsic limitation of the RDW process. In particular, the dominant low-frequency contributions are consistent with environmental perturbations and interferometer instabilities, as identified in the frequency noise analysis. Additionally, the oscillator in this work is operated without active stabilization, which can further increase the measured phase noise if the beat signal mixing process does not fully cancel the $f_0(t)$ contribution. Taken together, these results show that the measured differential phase noise is dominated by low-frequency technical fluctuations and therefore provides a conservative upper bound on any intrinsic phase noise introduced during RDW generation.

\section{Conclusion}

We have directly characterized the phase coherence of RDW emission at 260~nm using a UV f-4f interferometer. The measured f-4f beat exhibits the expected $3f_0(t)$ dependence, as confirmed by its correlated frequency shift with the simultaneously measured f-2f beat. By digitally mixing the f-4f signal with the frequency-tripled f-2f signal, we cancel the common $f_0(t)$ fluctuations. The resulting difference-frequency beat has a FWHM linewidth of 7.8~kHz over a 200~ms measurement period. The integrated RMS phase remains below 1~rad over the 1–30~kHz frequency range.

The residual noise is concentrated at low frequencies and includes discrete features consistent with technical noise arising from differential interferometer instabilities. Because the measured signal also contains noise from the f-2f nonlinear broadening process and both interferometers, the measured linewidth and phase noise provide conservative upper bounds on the excess noise introduced during RDW generation. A collinear or actively stabilized interferometer, together with improved beat note SNR and stabilization of the driving comb, could further reduce the technical noise and establish a tighter bound on the intrinsic RDW contribution. More broadly, the combined f-4f and f-2f approach provides a direct method for testing RDW coherence across different wavelengths, gases, driving conditions, and repetition rates. The present results at 2~MHz support RDW emission as a promising route to phase-coherent UV frequency combs.


\begin{backmatter}
\bmsection{Funding} This research was funded by the Max Planck-UBC-UTokyo Centre for Quantum Materials, the Canada First Excellence Research Fund in Quantum Materials and Future Technologies Program, the Gordon and Betty Moore Foundation's EPiQS Initiative Grant No. GBMF4779 to D.J.J., the Natural Sciences and Engineering Research Council of Canada Discovery Grants program, the Canada Foundation for Innovation and the British Columbia Knowledge Development Fund.

\bmsection{Acknowledgment}
We thank John Travers and Federico Belli for help on early stages of this work.
\smallskip

\bmsection{Disclosures} The authors declare no conflicts of interest.

\bmsection{Data availability} Data underlying the results presented in this paper are not publicly available at this time but may be obtained from the authors upon reasonable request.

\bigskip

\end{backmatter}

\bibliography{dwe_bibliography2}

\clearpage
\twocolumn[
\begin{center}
    {\LARGE\bfseries Supplementary Information\par}
    \vspace{0.5cm}
\end{center}]

\setcounter{section}{0}
\renewcommand{\thesection}{S\arabic{section}}
\setcounter{figure}{0}
\renewcommand{\thefigure}{S\arabic{figure}}
\setcounter{equation}{0}
\renewcommand{\theequation}{S\arabic{equation}}

\section{Comb equations}

\label{sec:supp_s1}
We can write the frequency of the $n$th comb element as 
\begin{equation}
\label{eq:oscillator_comb}
    f_n(t) = nf_{\mathrm{rep}}(t) + f_0(t),
\end{equation}
where $f_{\mathrm{rep}}$ is the repetition rate and $f_0$ is the carrier-envelope offset frequency. Here we have explicitly added time dependence to each quantity to indicate that these parameters have associated frequency noise.

Since the f-2f interferometer is seeded directly from the oscillator, We use Eq.~\ref{eq:oscillator_comb} to derive the f-2f interferometer beat signal. The $m$th element of the octave-spanning spectrum is given by
\begin{equation}
    f_m(t) = mf_{\mathrm{rep}}(t) + f_0(t) + f_{\mathrm{NL}}(t),
\end{equation}
where $f_{\mathrm{NL}}(t)$ is the frequency noise imparted by the supercontinuum generation process. The frequency doubled portion of the spectrum of comb element $n$ is written as
\begin{equation}
    2f_n(t) = 2nf_{\mathrm{rep}}(t) + 2f_0(t),
\end{equation}
assuming that the SHG process imparts no additional frequency noise to the comb. The heterodyne beat signal between the two arms is then given by
\begin{equation}
\label{eq:f2f_verbose}
\begin{split}
    f_{f-2f}(t) &= 2f_n(t) - f_m(t) \\
    &= (2n-m)f_{\mathrm{rep}}(t) + f_0(t) - f_{\mathrm{NL}}(t) + f_{\mathrm{int}}(t),
\end{split}
\end{equation}
where the additional term $f_{\mathrm{int}}(t)$ arises from path length instabilities in the interferometer itself. Under the condition that $2n = m$, Eq.~\ref{eq:f2f_verbose} reduces to
\begin{equation}
\label{eq:f2f_beat}
    f_{f-2f}(t) = f_0(t) - f_{\mathrm{NL}}(t) +  f_{\mathrm{int}}(t).
\end{equation}

The f-4f beat note frequency can be derived in a similar manner, with the additional complication that pulse picking modifies the oscillator comb. After (synchronous) pulse picking \cite{devriesAcoustoopticPulsePicking2015}, the comb element equation becomes
\begin{equation}
\label{eq:pulse_picked_comb}
    f_{n'}(t) = n'\frac{f_{\mathrm{rep}}(t)}{N} + f_0(t) + f_{AOM}(t),
\end{equation}
where $N$ is the repetition rate division factor and $f_{AOM}(t)$ is the RF carrier frequency driving the AOM. In our case the AOM driving frequency is derived from the 8th harmonic of the oscillator repetition rate and thus Eq.~\ref{eq:pulse_picked_comb} can be rewritten as
\begin{equation}
\label{eq:pulse_picked_comb_2}
    f_{n'}(t) = [n' + 8N]\frac{f_{\mathrm{rep}}(t)}{N} + f_0(t).
\end{equation}
From here, the RDW emission arm of the interferometer is given by
\begin{equation}
    f_{m'}(t) = [m' + 8N]\frac{f_{\mathrm{rep}}(t)}{N} + f_0(t) + f_{\mathrm{RDW}}(t),
\end{equation}
with $f_{\mathrm{RDW}}(t)$ being the additional frequency noise imparted by the RDW emission process. The fourth harmonic arm is given by
\begin{equation}
    4f_{n'}(t) = 4[n' + 8N]\frac{f_{\mathrm{rep}}(t)}{N} + 4f_0(t),
\end{equation}
and the resulting beat note frequency after mixing is
\begin{equation}
\begin{split}
    f_{f-4f}(t) &= 4f_{n'}(t) - f_{m'}(t) \\
    &= [4n' - m' + 24N]\frac{f_{\mathrm{rep}}(t)}{N} + 3f_0(t) - f_{\mathrm{RDW}}(t) + f_{\mathrm{int}}'(t),
\end{split}
\end{equation}
with $f_{\mathrm{int}}'(t)$ indicating that this separate interferometer may have its own individual instabilities. Again under the condition that the repetition rate term cancels ($4n' - m' + 24N = 0$), the equation simplifies to
\begin{equation}
\label{eq:f4f_beat}
    f_{f-4f}(t) = 3f_0(t) - f_{\mathrm{RDW}}(t) +  f_{\mathrm{int}}'(t).
\end{equation}

The resulting difference frequency component of the mixed beat note signal is given by
\begin{equation}
\label{eq:mixed_beat_note_freq}
\begin{split}
    f_{f-4f}(t) &- 3f_{f-2f}(t) = \\
    &\left([4n' - m' + 24N]\frac{f_{\mathrm{rep}}(t)}{N} + 3f_0(t) - f_{\mathrm{RDW}}(t) + f_{\mathrm{int}}'(t) \right) \\
    &- 3\bigg( [2n-m]f_{\mathrm{rep}}(t) + f_0(t) - f_{\mathrm{NL}}(t) + f_{\mathrm{int}}(t) \bigg)
\end{split}
\end{equation}

Assuming $4n' - m' + 24N = 0$ and $2n=m$ such that all of the repetition rate contributions cancel out Eq.~\ref{eq:mixed_beat_note_freq} becomes
\begin{equation}
\begin{split}
    f_{f-4f}(t) - 3f_{f-2f}(t) &= 
    \bigg(3f_0(t) - f_{\mathrm{RDW}}(t) + f_{\mathrm{int}}'(t) \bigg) \\
    &~~~~~- 3\bigg(f_0(t) - f_{\mathrm{NL}}(t) +  f_{\mathrm{int}}(t) \bigg) \\
    &= 3f_{\mathrm{NL}}(t) - f_{\mathrm{RDW}}(t) - 3 f_{\mathrm{int}}(t) + f_{\mathrm{int}}'(t).
\end{split}
\end{equation}

\section{Reconstructions of beat notes from frequency noise spectra}
\label{sec:supp_s2}

To validate the frequency noise extraction, we reconstructed the beat note lineshapes from their frequency noise spectra and compared them with the RF power spectral densities calculated directly from the same time-domain signals. Before extracting the instantaneous phase, each beat note component was isolated using a sixth-order hyper-Gaussian filter with amplitude transfer function
\begin{equation}
H(\nu)
=
\exp\left[
-\left(\frac{\nu-\nu_c}{\sigma}\right)^6
\right],
\label{eq:hypergaussian_filter}
\end{equation}
where $\nu_c$ is the filter center frequency and $\sigma$ is its scale parameter. The corresponding power transfer function is $\lvert H(\nu)\rvert^2$, with a full width at half-maximum of
\begin{equation}
\sigma_{\mathrm{FWHM}} =
2\left(\frac{\ln 2}{2}\right)^{1/6}\sigma
\approx 1.68\sigma.
\end{equation}
The values of $\sigma$ quoted below therefore refer to the scale parameter of the hyper-Gaussian filter rather than its full width at half-maximum.

For stationary Gaussian frequency fluctuations, the normalized autocorrelation function can be calculated from the single-sided frequency noise power spectral density according to \cite{didomenicoSimpleApproachRelation2010}
\begin{equation}
\label{eq:freq_noise_ac}
\Gamma(\tau)
=
\exp\left[
-2\int_0^\infty
S_{\Delta f}(f)
\frac{\sin^2(\pi f\tau)}{f^2}
\,\mathrm{d}f
\right].
\end{equation}
The corresponding RF lineshape is then obtained using the Wiener--Khinchin theorem,
\begin{equation}
\label{eq:rf_spectrum_from_ac}
S_{\mathrm{rec}}(\nu)
\propto
\int_{-\infty}^{\infty}
\Gamma(\tau)e^{-i2\pi\nu\tau}
\,\mathrm{d}\tau,
\end{equation} 
where $f$ is the Fourier frequency of the frequency noise spectrum and $\nu$ is the frequency offset from the beat note center. The reconstruction was evaluated by integrating the measured frequency noise up to 1~MHz. Increasing the integration range or numerical sampling did not appreciably change the reconstructed lineshapes. Because the instantaneous phase was extracted after digitally filtering each beat note, the reconstruction depends to some extent on the filter scale parameter $\sigma$. The values used here were chosen to isolate the desired beat note components without visibly truncating their central lineshapes, and the extracted frequency noise spectra were stable against moderate variations in $\sigma$.

\begin{figure}[ht]
\centering
\includegraphics[width=\linewidth]{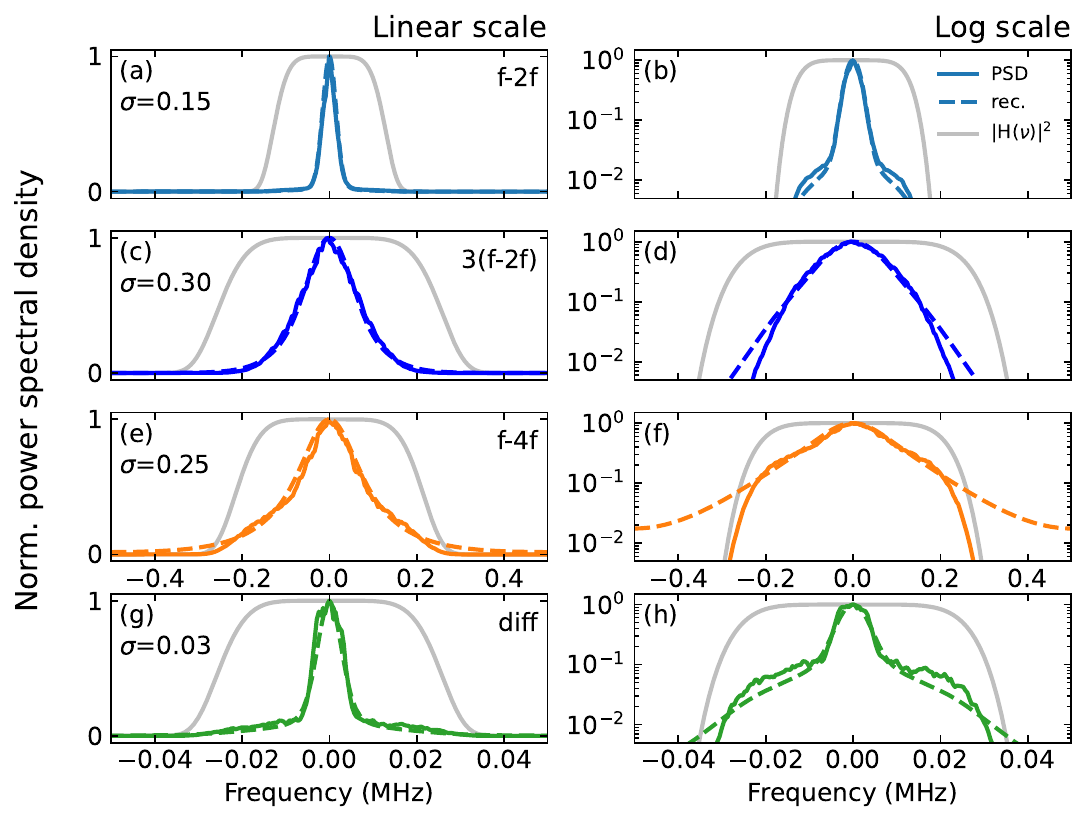}
\caption{Comparison of the beat note power spectral (PSDs) calculated directly from the filtered time-domain beat signals (solid lines) with the lineshapes reconstructed from their frequency noise spectra (dashed lines). The power transfer functions $\lvert H(\nu)\rvert^2$ of the sixth-order hyper-Gaussian filters are shown in gray. The reconstructed spectra closely reproduce the central lineshapes and relative linewidths of the directly calculated RF spectra, supporting the frequency noise extraction used in the main text.}
\label{fig:beat_windows_and_reconstructions}
\end{figure}

Figure~\ref{fig:beat_windows_and_reconstructions} compares the beat note power spectral densities calculated directly from the filtered time-domain signals with the lineshapes reconstructed from their frequency noise spectra. The left column shows the f-2f, frequency-tripled f-2f, f-4f, and mixed difference-frequency beat signals on a linear scale, while the right column shows the same spectra on a logarithmic scale. The directly calculated spectra are shown as solid coloured lines and the reconstructed lineshapes as coloured dashed lines. The filter power transfer functions $\lvert H(\nu)\rvert^2$ are shown in gray, with the corresponding values of $\sigma$ indicated in the left column. The reconstructions closely reproduce the central lineshapes and relative linewidths of the directly calculated RF spectra throughout the signal-processing chain. Some differences remain in the spectral pedestals, particularly for the lower-signal-to-noise f-4f beat. Nevertheless, the overall agreement provides a self-consistency check demonstrating that the extracted frequency noise spectra capture the fluctuations responsible for the observed beat note lineshapes.

\end{document}